\documentstyle[12pt]{article}
\begin{document}\baselineskip=18pt
\def\be{\begin{equation}}
\def\ee{\end{equation}}
\def\bearst{\begin{eqnarray*}}
\def\eearst{\end{eqnarray*}}
\def\peleven{\parbox{11cm}}
\def\peffec{\peight{\bearst\eearst}\hfill\peleven}
\def\pspace{\peight{\bearst\eearst}\hfill}
\def\ptwelve{\parbox{12cm}}
\def\peight{\parbox{8mm}}
\def\bear{\begin{eqnarray}}
\def\eear{\end{eqnarray}}
\def\E{{\rm e}}
\input epsf.tex
\newcommand{\slp}{\raise.15ex\hbox{$/$}\kern-.57em\hbox{$\partial$}}
\newcommand{\sla}{\raise.15ex\hbox{$/$}\kern-.57em\hbox{$a$}}
\newcommand{\slA}{\raise.15ex\hbox{$/$}\kern-.57em\hbox{$A$}}
\newcommand{\slB}{\raise.15ex\hbox{$/$}\kern-.57em\hbox{$B$}}
\newcommand{\slD}{\raise.15ex\hbox{$/$}\kern-.57em\hbox{$D$}}
\newcommand{\slb}{\raise.15ex\hbox{$/$}\kern-.57em\hbox{$b$}}
\newcommand{\slW}{\raise.15ex\hbox{$/$}\kern-.57em\hbox{$W$}}
\font\grrm=cmbx10 scaled 1200
\font\vb=cmbx10 scaled 1440
\font\bigcal=cmsy10 scaled 1200
\def\eightpoint{\def\rm{\fam0\eightrm}}
\def\flex{\raise 6pt\hbox{$\leftrightarrow $}\! \! \! \! \! \! }
\def\tr{ \mathop{\rm tr}}
\def\atanh{\mathop{\rm atanh}}
\def\Tr{\mathop{\rm Tr}}
\def\dal{\Box} 
\def\Natural{\hbox{\hskip 1.5pt\hbox to 0pt{\hskip -2pt I\hss}N}}
\def\Integer{\>\hbox{{\sf Z}} \hskip -0.82em \hbox{{\sf Z}}\,}
\def\Rational{\hbox{\hbox to 0pt{\hskip 2.7pt \vrule height 6.5pt
                                  depth -0.2pt width 0.8pt \hss}Q}}
\def\Real{\hbox{\hskip 1.5pt\hbox to 0pt{\hskip -2pt I\hss}R}}
\def\Complex{\hbox{\hbox to 0pt{\hskip 2.7pt \vrule height 6.5pt
                                  depth -0.2pt width 0.8pt \hss}C}}
\def \ln {{\rm ln}\, }
\def \cotg {\rm cotg }
\vskip 1cm
\begin{tabbing}
\hskip 11.5 cm \= \\
\>hep-th/9704176\\
\>April 1997
\end{tabbing}
\vskip 1cm
\begin{center}
{\Large\bf Screening in three-dimensional QED}
\vskip 1.2cm
{\large \bf E. Abdalla$^a$ and 
R. Banerjee$^b$\footnote{On leave of absence from\\
S.N. Bose National Centre for Basic Sciences\\
Block JD, Sector III, Salt Lake City, Calcutta 700.091 India}}\\
\vskip 0.4cm
{{\it $^a$Instituto de F\'\i sica-USP,\\ C.P.66318, 05315-970
S. Paulo, Brazil}\\
\vskip 0.4cm
{\it $^b$Instituto de F\'\i sica da Universidade Federal do Rio de Janeiro\\
CP 68.528, 21.945-970 Rio de Janeiro, Brazil}\\
$^a$ eabdalla@fma1.if.usp.br\\
$^b$ rabin@if.ufrj.br}
\end{center}
\abstract 

We compute the quark--antiquark potential in three dimensional
massive Quantum Electrodynamics. The result indicates that screening
prevails for large quark masses, contrary to the classical expectations.
The classical result is reproduced for small separation of the quarks.

\vfill\eject

A proper study of the problem of screening  and confinement is of
considerable importance in our understanding of gauge theories.
To avoid the complexities of four dimensions these studies are usually 
confined to lower dimensions. 
In this framework, a deep physical interpretation has been achieved. Indeed, 
in two--dimensional QED\cite{klaus-heinz}, one obtains screening in the 
massless case, but confinement in the massive quark case, realizing the 
expected picture. 

For QCD in two dimensions Gross et al \cite{gross} were the first to discuss
the subject. If dynamical fermions and test charges are in different
representations, they find screening or confinement in some particular cases
depending on whether the fermion is massless or massive.
A similar conclusion in an identical setting has been
arrived at for the massless case in \cite{frso}. If, on the other hand, all
fermions are in the fundamental representation, then screening
prevails independently of the quark mass \cite{amz}.

General inquires in two dimensional gauge theories have been performed
recently by several authors \cite{semenov}, concerning the $\theta$ 
vacuum structure, screening, confinement and chiral condensates. In 
three dimensionons  related questions were studied in \cite{semqed3}.

It is thus important and instructive to verify how far such issues are just
low dimensional unphysical features, or part of the theoretical structure
of gauge theories. Usually, the probe of confinement comes from the
Wilson criteria, computing the Wilson loop and checking whether it behaves 
as the area or the perimeter of the loop, for large loops\cite{wilson}.
This is also the approach followed by \cite{gross}. Here we follow an 
alternative route based on the direct calculation of the quark-antiquark 
potential. To perform the computations we shall take recourse to 
bosonisation. This is a well known technique in two-dimensional
space-time\cite{coleman-mandelstam} which has been well illustrated 
in getting the  Schwinger
terms in the current algebra in fermionic field theories, and 
in order to study the problem of screening and confinement in 
QED$_2$\cite{klaus-heinz} as well as in QCD$_2$\cite{amz}.  
This is possible because one is led to effective actions 
which contain quantum effects already at the classical level.

The familiar ideas of two dimensional bosonisation have been recently
extended to higher dimensions and, in particular, a bosonised form for massive 
QED$_3$ has been 
developed\cite{boso3}. We use this formulation to investigate the phenomenon 
of screening and confinement in this theory by explicitly computing
the quark-antiquark potential. The result shows
that contrary to the classical expectation,  there is
screening for large quark mass. However the classical result is reproduced
for a small separation of the quarks.

The partition function of three dimensional massive QED
in the covariant gauge, in the presence of an external source $J^\mu$, 
is given by
\begin{eqnarray}
Z &= &\int d[\psi, \bar{\psi},
A_{\mu}]\delta(\partial_{\mu} A^{\mu}){\em exp}\
i\int d^3 x(\bar{\psi}(i\partial\!\!\!/\, - m -  e
A\!\!\!\!/\,)\psi \nonumber \\
&  &- \frac{1}{4} F^{2}_{\mu\nu} + A_\mu J^\mu)\label{partition-function}
\end{eqnarray}
where $F_{\mu\nu}$ is the field tensor, $F_{\mu\nu}=\partial_\mu A_\nu -
\partial_\nu A_\mu $.

The bosonised version of the above defined action in the large mass
limit is given by the expression\cite{boso3}
\begin{eqnarray}
Z &= &\int dA_{\mu}\delta(\partial_{\mu}A^{\mu})exp\ i \int d^{3}x[-
\frac{ e^{2}}{8\pi}
\epsilon_{\mu\nu \rho}A^{\mu}\partial^{\nu}A^{ \rho} \nonumber \\
  &  &+(\frac{ e^{2}}{24\pi m} - \frac{1}{4}) F^{2}_{\mu\nu} +
A_\mu J^\mu  +.....]\label{bosoaction}
\end{eqnarray}
where terms up to order $1/m$ have been retained in the computation
of the effective action as a power series in the inverse quark mass. 
This result is just the partition function of
the Maxwell-Chern-Simons\cite{djt} theory in the covariant gauge.

We now compute the potential as being the difference between the 
Hamiltonian with and without a pair of static external charges separated by 
a distance $L$, so that
\bear
V(L)&=& H_q-H_0 =  -(L_q-L_0)\nonumber\\
&=& -q\int d^2x A_\mu \delta^{\mu 0}
\lbrack \delta (x^1+L/2)\delta (x^2) -\delta (x^1-L/2)\delta (x^2)
\rbrack \nonumber\\
&=& -q\lbrack A_0(x^1=-L/2,x^2=0)- A_0(x^1=L/2,x^2=0)\rbrack\; .
\label{v-as-extsource}
\eear
where we have integrated over the two space components in order to
find the potential, and considered the source as corresponding to two
fixed charges of magnitude $q$ located at the points defined by the 
respective delta functions. Note that $L_q (L_0)$ denote the Lagrangeans in
the presence (absence) of the charges.

We now consider the equations of motion associated with the Lagrangean 
defined in \ref{bosoaction}. The field equation in the covariant gauge reads
\be
-{ e^2\over 4\pi} \epsilon _{\mu\nu\rho}\partial^\nu A^\rho
+(1-{ e^2\over 6\pi m})\Box A_\mu +J_\mu =0
\label{a-eq-motion}
\ee

Defining the curl of $A_\mu$ as
\be
{\cal A}_\mu = -\epsilon_{\mu\nu\beta}\partial^\nu A^\beta\label{curl}
\ee
the equation of motion can be expressed as
\be
\lbrack \Box +m_A^2\rbrack {\cal A}_\mu =-{1\over 1-{ e^2\over 6\pi m}}
\epsilon_{\mu\nu\beta}\partial^\beta J^\nu - {{ e^2\over 4\pi}\over
\lbrace 1-{ e^2\over 6\pi m}\rbrace^2 }J_\mu\label{eq-motion-curl-a}
\ee
where $m_A={{ e^2\over 4\pi}\over
\lbrace 1-{ e^2\over 6\pi m}\rbrace } $. In the absence of sources it
reproduces the familiar massive mode of Maxwell-Chern-Simons theory\cite{djt}.
From (\ref{v-as-extsource}) it is seen that an expression for $A_0$
is required to calculate the potential. This is given in terms
of the curl (\ref{curl}) by
\be {\cal A}_2=-\partial_1A_0\label{a2a0}
\ee

The time independent solution for ${\cal A}_2$ corresponding to the 
sources describing static quarks 
can be obtained from (\ref{eq-motion-curl-a}). Using this result with
(\ref{a2a0}) finally yields
\bear
A_0(x) &=&-\int d^2 y \Delta (x-y,m_A) {1\over  1-{ e^2\over 6\pi m}}
J_0(y)\nonumber\\
&=&\!\! -{q\over  1\!-\!{ e^2\over 6\pi m}}\lbrack 
\Delta(x^1+\frac L 2,x^2;m_{_A})
\!-\!\Delta(x^1-\frac L 2,x^2;m_{_A})\rbrack \label{sol-static-quarks}
\eear
where $\Delta(x;m_A)$ is the Euclidean Feynman propagator in two dimensions,
since we are dealing with the time independent Greens function in
(\ref{eq-motion-curl-a}). It is given by the modified Bessel function,
\be
\Delta(x; M) ={1\over 2\pi} K_0 (M\sqrt {(x^1)^2+(x^2)^2} )\label{bessel}
\ee
The above solution (\ref{sol-static-quarks}) is defined up to an $x^2$ 
dependent constant, which will be overlooked in what follows, since it 
will not affect our results.

The potential is now found from (\ref{v-as-extsource}) and 
\ref{sol-static-quarks}, reading
\be
V(L)=2{q^2\over 1-{ e^2\over 6\pi m}}\lbrack \Delta (0,0;m_A)-
\Delta(L,0;m_A)\rbrack\label{v-of-l}
\ee
At this point we disregard the constant term, as discussed before, arriving
at the main result of this work,
\be
V(L)=-{1\over \pi}{q^2\over 1-{ e^2\over 6\pi m}}K_0(m_AL)
\equiv -{q_{ren}^2\over \pi}K_0(m_AL)
\label{v-as-bessel}
\ee
The behaviour of this potential energy should be compared with the classical
expectation, that is
\be
V^{class}(L)={q^2\over\pi}\ln \left(m_A L\right)\label{v-classical}
\ee
where we arbitrarily choose the massive parameter. The classical result
in the ultraviolet regime corresponds to a renormalisation of the strength,
but in the infrared, a logarithmic growth is expected, which
would signalize confinement of the external quarks. However,
as in the case of two--dimensional QCD the quantum result 
(\ref{v-as-bessel}) indicates screening,
due to the Chern-Simons term in the action, which  induces
a mass for the gauge field. Interestingly, in the limit
$L\to 0$ (i.e. when the quarks are close), the leading term in
the expression
(\ref{v-as-bessel}) for heavy quarks reduces to the classical result
(\ref{v-classical}).
In the two--dimensional case, this is also true, e.g. the classical potential
for short separations is given by the expression
\be
V^{class}(L)={e^2\over 2}L
\ee
which corresponds to the short separation limit (for heavy quarks)
of the full quantum result\cite{klaus-heinz,amz}.

In the three dimensional case, we drew the diagrams corresponding to the
classical and to the quantum results superposed in figure [1] for comparison.

We see therefore that the conclusions
obtained in two--dimensional space--time are valid in the
three dimensional case, strengthening them, and providing
further reality to the results. 
The utility of the bosonisation methods in the present context has 
been clearly illuminated. These methods
prove to be of greater effectiveness in obtaining physical results,
especially due to the reason that the bosonised version contains quantum 
corrections at the classical level.

To put our work in the proper perspective we recall that screening effects
in the Maxwell-Chern-Simons theory may have been known \cite{djt} but
the observation that there exists
a direct connection between fermionic three dimensional QED and 
the Maxwell-Chern-Simons theory leading to similar effects in the
former is new, as shown in our work.
Moreover we gave an explicit calculation for the quark-antiquark
potential for large values of the quark mass which quantitatively 
illuminated the screening phenomenon due to the
quantum effects. Indeed the departure of the quantum result from the
classical expression was clearly illustrated (see figure). 
It is also appropriate to mention that the results obtained here were based on
bosonisation in the large mass limit using the quadratic approximation. 
It is of course possible to extend the bosonisation scheme to arbitrary
mass \cite{bm} or to go beyond the quadratic approximation. It would then
be interesting to see whether the screening phase persists or is modified
under these circumstances. As a concluding
remark we mention that screening effects in QED$_3$ have not been
investigated either using bosonisation or any other method.

\begin{figure}[b]
\begin{center}
\leavevmode
\bearst
\epsfxsize= 9truecm{\epsfbox{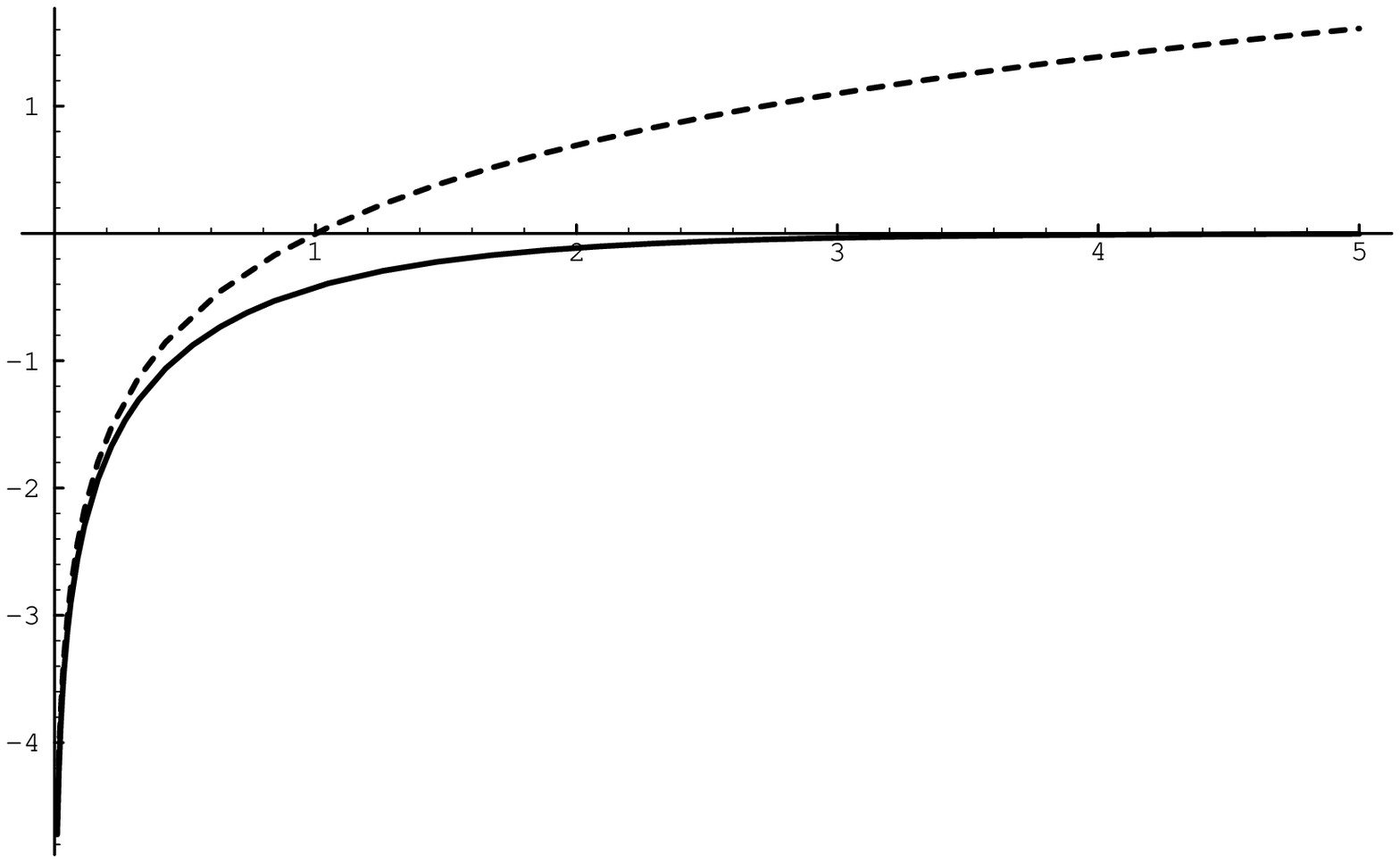}} 
\eearst 
\vskip .5cm
\caption{{\it Classical (dashed line) and quantum (continuous line) 
effective inter--quark potentials are sketched as a function of their
separation. The classical potential grows with inter--quark separation
(confinement) while for the quantum theory, the potential
tends asymptotically to zero.}}
\end{center}
\end{figure}

\vskip 1cm

{\bf Acknowledgements}: this work  has been partially supported
by Conselho Nacional de Desenvolvimento Cient\'\i fico e Tecnol\'ogico,
CNPq, Brazil, and Funda\c c\~ao de Amparo \`a Pesquisa do Estado de
S\~ao Paulo (FAPESP), S\~ao Paulo, Brazil.
R.B. thanks the physics department of the UFRJ and USP for the kind 
hospitality.


\end{document}